\documentclass[10pt,conference]{IEEEtran}
\IEEEoverridecommandlockouts
\usepackage{cite}
\usepackage{amsmath,amssymb,amsfonts}
\usepackage{algorithmic}
\usepackage{graphicx}
\usepackage{textcomp}
\usepackage{xcolor}

\usepackage[nameinlink,noabbrev,capitalise]{cleveref}

\usepackage{epigraph}
\usepackage{todonotes}
\usepackage{csquotes}

\def\BibTeX{{\rm B\kern-.05em{\sc i\kern-.025em b}\kern-.08em
    T\kern-.1667em\lower.7ex\hbox{E}\kern-.125emX}}
\begin{document}

%
%

\title{The Mind Is a Powerful Place:\\ How Showing Code Comprehensibility Metrics Influences Code Understanding}

\author{
\IEEEauthorblockN{Marvin Wyrich\IEEEauthorrefmark{1},
Andreas Preikschat\IEEEauthorrefmark{2}, Daniel Graziotin\IEEEauthorrefmark{1} and
Stefan Wagner\IEEEauthorrefmark{1}}
\IEEEauthorblockA{
Institute of Software Engineering,
University of Stuttgart\\
Stuttgart, Germany\\
\IEEEauthorrefmark{1}\{firstname.lastname\}@iste.uni-stuttgart.de,
\IEEEauthorrefmark{2}andreaspreikschat@posteo.de}
}


\maketitle
\IEEEpeerreviewmaketitle

%
%

\begin{abstract}
Static code analysis tools and integrated development environments present developers with quality-related software metrics, some of which describe the understandability of source code. Software metrics influence overarching strategic decisions that impact the future of companies and the prioritization of everyday software development tasks.
Several software metrics, however, lack in validation: we just choose to trust that they reflect what they are supposed to measure. Some of them were even shown to not measure the quality aspects they intend to measure. Yet, they influence us through biases in our cognitive-driven actions. In particular, they might anchor us in our decisions. Whether the anchoring effect exists with software metrics has not been studied yet.
We conducted a randomized and double-blind experiment to investigate the extent to which a displayed metric value for source code comprehensibility anchors developers in their subjective rating of source code comprehensibility, whether performance is affected by the anchoring effect when working on comprehension tasks, and which individual characteristics might play a role in the anchoring effect.
We found that the displayed value of a comprehensibility metric has a significant and large anchoring effect on a developer's code comprehensibility rating. The effect does not seem to affect the time or correctness when working on comprehension questions related to the code snippets under study.
Since the anchoring effect is one of the most robust cognitive biases, and we have limited understanding of the consequences of the demonstrated manipulation of developers by non-validated metrics, we call for an increased awareness of the responsibility in code quality reporting and for corresponding tools to be based on scientific evidence. \footnote{\copyright 2021 IEEE. Personal use of this material is permitted. Permission from IEEE must be obtained for all other uses, in any current or future media, including reprinting/republishing this material for advertising or promotional purposes, creating new collective works, for resale or redistribution to servers or lists, or reuse of any copyrighted component of this work in other works. Postprint, after review. \newline To appear in: Proceedings of the 43rd International Conference on Software Engineering (ICSE '21), Madrid, Spain, 12 pages}
\end{abstract}


%
%

\section{Introduction\label{sec:intro}}
Software developers spend more than 50\% of their time on activities related to program understanding~\cite{Minelli:2015:LastSummer,Xia:2018:Measuring}.
Development teams strive to make their code as understandable as possible---refactoring source code to make it more understandable is a central part of agile software methodologies~\cite{alshayeb2009empirical,beck2000extreme}---and base their activities on the results of static code analysis tools to identify areas of code that are still difficult to understand.

Most of the metrics reported by such tools are either not validated~\cite{Nilson:2019:Validated} to the point that some are empirically demonstrated to not measure what they are assumed to measure~\cite{Scalabrino:2019:Automatically}.
The latter issue seems to be especially prevalent in the field of code comprehensibility~\cite{Scalabrino:2019:Automatically,Baron:2020:CogComplexity}. In other words: several metrics do not reflect what they are supposed to measure. Yet, they are considered when making decisions and change course of what developers think of their source code. What is being shown, or told, to people influences what they think.

\epigraph{\textit{The mind is a powerful place / And what you feed it can affect you in a powerful way}}{---NF, \textit{The Search} (song). 2019.}

Feeding the mind with a belief influences it and causes changes that might go beyond the mind itself. Crum and Langer~\cite{Crum:2007:Mindset} divided a sample of room attendants at different hotels into two groups. To the first one only, the researchers presented the supposed positive effects of work-related physical activities on their health. After four weeks, the informed group felt that they received significantly more exercise than the second group. Not just that: Weight, blood pressure, and body fat of the informed group significantly decreased compared to the non-informed one---without any detected change in workload, outside work physical activity, or eating habits. If the consequence of a treatment is not attributed to the treatment itself, but to pure beliefs and expectations of its effectiveness, we call it the \textit{placebo effect}~\cite{Shapiro:1968:SemanticsPlacebo}.
This effect occurs in many ways. Placebos are administered in clinical trials in the form of sham drugs to distinguish the pharmaceutical effect of a drug from the placebo effect. The placebo effect can go beyond the subjective perception of an affected person and provide measurable effects. The room attendants were specifically manipulated by the researchers and a desirable effect was achieved.

Various contextual factors can also influence our reasoning and decision making, some of which make us deviate from beneficial results. Many of these factors are cognitive in nature~\cite{hilbert2012toward}. Software engineering is no exception. In a recent systematic mapping study on cognitive biases in software engineering, 65 articles were identified that provide evidence for the presence of cognitive biases of at least eight different categories~\cite{Mohanani:2018:Cognitive}. Negative consequences of such biases are, for example, overly optimistic effort estimates or insufficient software modifications.

A cognitive bias which is related to estimates and adjustments is the \textit{anchoring effect}, introduced by Tversky and Kahneman~\cite{Tversky:1974:Anchoring}. The anchoring effect means that an initial value is insufficiently adjusted so that \enquote{different starting points yield different estimates, which are biased toward the initial values}~\cite{Tversky:1974:Anchoring}. The anchoring effect is one of the most robust cognitive biases~\cite{Furnham:2011:ReviewAnchoring}.

In our study, which we align with the field of quantitative behavioral software engineering~\cite{lenberg2015behavioral,graziotin2020psychometrics}, we investigate the anchoring effect in the context of source code comprehensibility\footnote{We use the terms comprehensibility and understandability interchangeably, in line with some of our peers, e.g.,~\cite{Scalabrino:2019:Automatically,Hofmeister:2017:ShorterIds}.} and software metrics.

We conducted a randomized, double-blind experiment. Participants were divided into two groups and were asked to work on source code comprehension tasks over code snippets. We showed the two groups a metric value that represents the understandability of the code snippets. One group saw a value that indicates an easy understandability of the source code. The other group saw one that indicates a hard understandability of the source code. Unbeknownst to the participants, tasks and snippets were the very same for both groups. Also, the metric is not real and placed there to anchor them.

We investigated the extent to which the shown metric anchors our participants in their understandability. As common in code comprehension experiments, we assessed the effect through the subjective evaluation of the participants~\cite{Oliveira:2020:Evaluating,Baron:2020:CogComplexity}. We also aimed to find out whether the manipulation leads to a placebo effect of the form that we measure differences in actual code comprehension performance, i.e., the time spent and correctness in answering comprehension questions. Finally, we explored an initial pool of individual characteristics (such as experience, personality, and mood) that are envisioned to play a role in the anchoring effect despite a very limited availability of studies~\cite{Furnham:2011:ReviewAnchoring}.

To these ends, we formulated three research questions, which are further framed later in section~\ref{sec:background} and in section~\ref{methods_variables}:

\begin{itemize}
\item \textbf{RQ1:} Does the value of a shown code comprehensibility metric influence subjective ratings of code comprehensibility?
\item \textbf{RQ2:} Does the value of a shown code comprehensibility metric influence the actual code understanding?
\item \textbf{RQ3:} To which extent do selected individual characteristics correlate with the deviation of the subjective rating from the shown metric value?
\end{itemize}

Understanding the consequences of displaying a metric value brings valuable practical and theoretical implications. First, showing metric values without proper validation can lead to the revision of already well understandable source code, which would unnecessarily take time and may also introduce new defects. Second, if the presence of a single metric value significantly influences a developer's rating, this would be a strong call to meticulously control program comprehension experiments for potential confounding variables that could bias a developer's judgement or to advise against this measurement method altogether.

%
%

\section{Background\label{sec:background}}
In this section we define the central constructs of this work, namely the placebo effect, the anchoring effect and source code comprehension, and place the work in the context of related literature. Given the limited availability of related work in the field of software engineering, we include related work in the background section, under each of the three topic-centered subsections that follow.

\subsection{Placebo Effect}

Probably one of the most quoted definitions of \textit{placebo} comes from Shapiro~\cite{Shapiro:1968:SemanticsPlacebo}, who studied the etymology and semantics of the word to provide a basis for an appropriate definition and to address the diversity of opinion about the meaning of the term.
The proposed definition is as follows:

\begin{displayquote}
A \textit{placebo} is defined as any therapy (or that component of any therapy) that is deliberately used for its nonspecific psychologic or psychophysiologic effect, or that is used for its presumed specific effect on a patient, symptom, or illness, but which unknown to therapist and patient is without specific activity for the condition being treated.~\cite[p.~682]{Shapiro:1968:SemanticsPlacebo}
\end{displayquote}

The \textit{placebo effect} is defined as \enquote{the nonspecific psychologic or psychophysiologic effect produced by a placebo}~\cite{Shapiro:1968:SemanticsPlacebo}.
The introduction of the term in medical literature was accompanied by \enquote{the widespread introduction of controlled methodology in the evaluation of treatment}~\cite{Shapiro:1968:SemanticsPlacebo} and it became standard to control for the placebo effect in clinical trials.
In this paper we follow Shapiro's definition.
We would like to point out that the placebo effect, however, is not limited to the medical context and can be applied to everyday aspects, as numerous studies have shown.

In a study on \textit{placebo sleep}~\cite{Draganich:2014:PlaceboSleep}, participants had to report their previous night's sleep quality.
One group was then told after a supposedly reliable measurement that their sleep quality was above average and the other group was informed that their sleep quality was below average.
The assigned sleep quality, but not the self-reported sleep quality, significantly predicted, among others, the auditory information processing speed of the participants.
The authors conclude that mindset can influence cognitive performance both positively and negatively~\cite{Draganich:2014:PlaceboSleep}.

Other studies show, for example, that smelling a supposedly creativity-enhancing odorant actually results in a creativity-enhancing effect~\cite{Rozenkrantz:2017:Creativity}, that non-invasive sham brain stimulation improves learning performance~\cite{Turi:2018:CognitivePlacebo}, and that different forms of placebos have an effect on the performance of athletes~\cite{Berdi:2011:PlaceboSport}.

Investigations of the placebo effect in software engineering research have rarely been conducted so far.
One recent study deals with the influence of a three-minute breathing exercise on the perceived effectiveness of stand-up meetings in agile project teams~\cite{Heijer:2017:Breath}.
A placebo group was added to compare the effect with a non-meditative form of relaxation, i.e. listening to classical music.
They conclude that the breathing exercise has an immediate positive impact on meetings in agile teams.
Another study~\cite{Jerffeson:2018:SBSE} investigates how the subjective evaluation of an automatically generated solution is positively influenced by involving the decision maker in the process but not considering their decisions at all.
They conducted a placebo-controlled study with 12 software engineering practitioners and found an increase of 68\% in the subjective evaluation of an automatically generated but supposedly decision influenced solution is due to a placebo effect.
We are not aware of any study investigating a potential placebo effect on performance in code understanding activities.

\subsection{Anchoring Effect}
In our behaviors, we act within a specific context. Such context provides us with cues, verbal suggestions, and social information that influence our expectations, appraisals and memories, which in turn influence our behavior and reported experiences~\cite{Wager:2015:Neuroscience}.
Consequently, parts of the placebo effect on subjective assessments are attributed to various forms of decision bias~\cite{Wager:2015:Neuroscience}.

A systematic mapping study on cognitive biases in software engineering highlights that the everyday life of a software engineer is also full of situations in which their decisions are subconsciously manipulated~\cite{Mohanani:2018:Cognitive}.
Mohanani et al. identified 65 articles in the context of software engineering that investigated 37 cognitive biases of at least eight different categories.
In the worst case, such bias leads to systematic deviations from optimal reasoning, such as overly optimistic effort estimates or insufficient software modifications~\cite{Mohanani:2018:Cognitive}.

The specific cognitive bias that we investigate is called \textit{anchoring effect}, which we defined in section~\ref{sec:intro}.
According to the aforementioned mapping study it is the most frequently investigated cognitive bias in the context of software engineering~\cite{Mohanani:2018:Cognitive}.
For example, one study used SQL queries as an anchor for query formulation tasks.
They found that while subjects complete the tasks more quickly when modifying a query instead of writing it from scratch, accuracy decreases and overconfidence in the results increases~\cite{Allen:2006:LittleHelp}.
Another example where anchoring plays a role is planning poker.
In planning poker, it is considered as positive that all effort estimates remain initially hidden from view, so that no one is anchored in their initial estimate by the estimates of their colleagues~\cite{Haugen:2006:Poker}.

The anchor would not even have to be relevant for the estimate~\cite{Furnham:2011:ReviewAnchoring} and could, for example, result from the previous turning of a wheel of fortune with numbers between 0 and 100~\cite{Tversky:1974:Anchoring}.
Since we are interested in the transfer of the anchoring effect to a realistic software engineering scenario, we have decided to display a code comprehensibility metric value.

Limited literature is available that has shown that individual characteristics of the participant may influence the strength of the anchoring effect.
Furnham and Boo argue in their literature review on the anchoring effect~\cite{Furnham:2011:ReviewAnchoring} that previous research \enquote{neglected
individual differences variables because people tend to look for a universal rule that would predict reactions or behaviour}.
Nevertheless, they could identify a total of 17 studies that considered the influence of experience, personality, mood, motivation and cognitive abilities.
We aim to contribute to this investigation in the context of RQ3 and explored the influence of experience, personality, happiness and dispositional optimism and pessimism on the anchoring effect.
All these constructs are also associated in literature with the placebo effect~\cite{Geers:2010:Dispositional,Morton:2009:Reproducibility,Wager:2015:Neuroscience,Pecina:2013:Personality}.
In~\ref{methods_materials} and~\ref{methods_variables} we describe the used questionnaires and how we have operationalized and measured the constructs.

\subsection{Source Code Comprehension/Understanding}
Source code understandability is defined as the extent to which ``code possesses the characteristic of understandability to the extent that its purpose is clear to the inspector''~\cite{Boehm:1976:Quantitative} and in this study, we particularly consider bottom-up comprehension, in which the programmer analyzes the source code line by line and from several lines, deduces \lq chunks\rq{} of higher abstraction and finally aggregates these chunks into high-level plans~\cite{OBrien:2004:BottomUp}.
Many factors impact the comprehensibility of source code.
For example, one study has shown that shorter identifiers take longer to comprehend~\cite{Hofmeister:2017:ShorterIds}, and another that a number of certain code patterns lead to an increased rate of misunderstanding~\cite{Gopstein:2017:UnderstandingMis}.

Static code analysis tools attempt to measure code understandability automatically to efficiently point out sections of the code that are difficult to comprehend and should therefore be refactored.
Not only are most of the metrics in static code analysis tools not validated~\cite{Nilson:2019:Validated}, but in addition there seems to be only one metric that is validated and positively correlates with measures of source code comprehensibility~\cite{Scalabrino:2019:Automatically,Baron:2020:CogComplexity}.

Scalabrino et al.~\cite{Scalabrino:2019:Automatically} investigated 121 metrics that would measure source code understandability.
Code snippets were evaluated with 63 developers and various proxy variables that are related to the time and correctness required to complete comprehension tasks.
According to their results, none of the investigated metrics showed a significant correlation with the measured source code comprehensibility.

A recent study~\cite{Baron:2020:CogComplexity} empirically evaluated the \textit{Cognitive Complexity}, a newly introduced metric that claims to measure source code understandability~\cite{Campbell:2018:CogComplexity}.
The metric evaluates code syntactically and assigns each method a calculated value on a ratio scale.
For Java methods, the authors suggest a threshold value of 15 above which a snippet should be refactored.
The authors of the evaluation study conclude that the metric is a reliable predictor of the required understanding time and the subjective comprehensibility rating of developers~\cite{Baron:2020:CogComplexity}.
Both aspects encouraged us to consider the metric when selecting code snippets for our study (see \ref{methods_materials}; they all had a value of 19).

Displaying non-validated metrics can lead to confusion and unnecessary effort due to improper prioritization of development efforts.
It becomes especially problematic if the displayed metric value actually has an influence on developers, even if only in their subjective perception of the code because this would mean that they are subconsciously manipulated and are very unlikely to resist this circumstance since the anchoring effect is one of the most robust cognitive processes~\cite{Furnham:2011:ReviewAnchoring}.

Finally, in experiments like ours, we do not intend to measure how understandable source code is, but the degree to which a participant has understood the given source code.
It is in the nature of our study to investigate selected influences on code understanding and not to compare variants of source code for their comprehensibility.
The most common measures for this purpose are time and correctness in processing comprehension questions, subjective ratings and physiological measurements such as eye tracking~\cite{Oliveira:2020:Evaluating,Baron:2020:CogComplexity}.
Through a literature review, Siegmund and Schumann~\cite{Siegmund:2015:Confounding} have compiled a list of confounding variables that can affect code comprehension.
Their catalog of confounding variables and control techniques is interesting not only because we have taken it into account for the design of our experiment but also because it reveals whether studies in the past have mentioned cognitive biases as confounding variables.
While we cannot identify in their list of \textit{individual} confounding parameters any that correspond to this, in the list of \textit{experimental} confounding parameters we found a handful that are related to cognitive biases.
One example is the Hawthorne effect~\cite{Roethlisberger:1939:Management, Mccarney:2007:Hawthorne}, which describes that participants in experiments would behave differently because they were observed.
With the present study we close a research gap and investigate whether the anchoring effect could add an entry to the catalog of confounding variables on code comprehension in the future.

%
%

\section{Methodology}

We follow the guidelines of Jedlitschka et al. on reporting experiments in software engineering~\cite{Jedlitschka:2008:Reporting}.

\subsection{Goals}

The goal of our study is to analyze the effect of showing a specific code comprehensibility metric on measures of a software engineer's code understanding to identify a potential cognitive bias and placebo effect.
To this end we formulated the three research questions given in the introduction.

\begin{figure*}[t]
\centering
\includegraphics[trim= 1cm 0 3cm 0, clip, width=\textwidth]{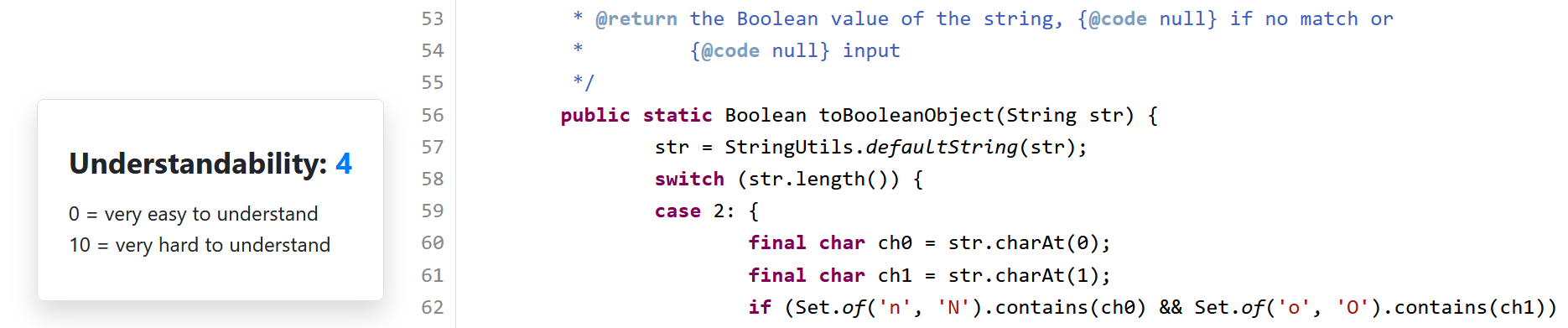}
\caption{Look and feel of the development environment that all participants used for the code comprehension tasks.}
\label{fig:code_environment}
\end{figure*}

\subsection{Participants}
\label{methods_participants}
We invited a convenience sample of students of a software engineering MSc study program in Germany to participate in the study.
According to recently proposed guidelines on sampling in software engineering~\cite{Baltes:2020:Sampling}, convenience sampling is \enquote{ideal for pilot studies and studying universal phenomena such as cognitive biases}.
Additionally, we limited participants to those with good knowledge of Java and German, as the study was conducted in German; in both aspects we relied on the self-assessment of the participants.
We see the sample properties of interest, namely enough experience to comprehend medium to hard to understand Java code, ensured by our sampling strategy so that the findings can be transferred to a population of experienced Java software engineers.
Limitations of our sampling strategy and their implications are discussed in \ref{sec:limitations}.

As part of their study duties, students had to participate in any study offered by the faculty. Students had the right to register for a study and then withdraw their participation at any time (including before the start) without consequences, with course organizers being unaware of their withdrawal. We reminded them about this during the informed consent phase, which included a partial design disclosure, health risks, privacy and ethical issues, and our contact details. Consent was obtained in written form.

Participants knew that we aimed to investigate factors that influence the understanding of source code, and that they would have to work on short methods written in Java and calculate the results for given input values. They were not aware of the metric manipulation.

\subsection{Tasks}
\label{methods_tasks}

Participants were shown three independent Java methods, one after the other.
For each code snippet, five input values were given, for which the participants were asked to specify the return values according to the Javadoc and to determine the actual return value.
Since we told our participants that there might be bugs in the code, they could not rely on the Javadoc comment and had to understand what the code actually does.
We consider the deviation of the documentation from the code and the inspection based on concrete values for the input parameters to be a realistic scenario.
Furthermore, the task is in line with the conceptual model that a developer in a maintenance scenario iteratively constructs and tests hypotheses about the functioning of the code during program understanding~\cite{Mayrhauser:1995:ProgramComp}.

Right after determining the return values, the participants were asked to rate the comprehensibility of the method on a scale of 0 (very easy) to 10 (very hard) and fill out questionnaires on their individual characteristics (details in section \ref{methods_procedure}).

\subsection{Experimental Materials}
\label{methods_materials}

\subsubsection{Environment}
The tasks were all solved on a laptop provided by us.
Code snippets were presented in a web environment specially developed for this study.
The look and feel of the web environment is based on the Eclipse IDE default look. Tooltips for variables and functions were displayed as typically expected in IDEs when hovering them. Syntax highlighting and line numbers were available. Selecting a variable highlights all occurrences of that variable. Next to the source code, the comprehensibility value of the method was displayed. A screenshot of the environment is shown in Fig.~\ref{fig:code_environment}.

\subsubsection{Code Snippets}

We used a total of five Java code snippets to conduct the study, two of which were used to introduce the study and explain the task, and the remaining three had to be understood by the participants. All participants received the same Java code snippets, regardless of treatment. Each code snippet consisted of exactly one class with exactly one method. The method was documented via Javadoc.

The three task-related snippets were taken from either the Apache Commons Lang or Apache Commons Collection project. We selected the snippets in a way that no uncommon prior knowledge on, e.g., frameworks, would be required to understand them. As a result, the code contained mostly primitive data types and the features of newer Java versions were avoided. The snippets were slightly modified, either to introduce a bug or to make sure that all task snippets have the same cognitive complexity, an indicator for the comprehensibility of the method, which is particularly reliable regarding the subjective rating of developers~\cite{Campbell:2018:CogComplexity,Baron:2020:CogComplexity}.
This allowed us to weigh the answers to the three tasks equally and limited potential confusion or loss of trust in the displayed metric if the same metric value was displayed (by design) but very different difficulties were perceived.
The three task snippets had a cognitive complexity score of 19, which is considered moderate to difficult to understand for Java methods.

\subsubsection{Comprehension Questions}

For each of the three tasks a participant was provided with a paper-based form which included five rows of a three-column table that had to be filled in.
The cells of the first column each contained a method call, for example \verb toBooleanObject("ofo") .
The other two columns had to be filled with the actual return value of the method and the expected return value according to the Javadoc.

\subsubsection{Questionnaires}

Participants had to fill out several questionnaires. Related constructs are detailed in the next section.

To assess happiness, we use the Scale of Positive and Negative Experiences (SPANE)~\cite{Diener:2010:SPANE} which quantifies the frequency of positive (SPANE-P) and negative (SPANE-N) affective experiences, and the happiness overall of our participants (SPANE-B). The questionnaire was successfully used (and fully described) in other studies of behavioral software engineering, e.g.,~\cite{graziotin2017unhappiness,graziotin2014happy}.
To appraise personality traits, we use the Big Five Inventory~\cite{Digman:1990:FiveFactor,McCrae:1992:FiveFactor}.
To measure dispositional optimism and pessimism, we use the Life Orientation Test (LOT-R)~\cite{Scheier:1994:LOTR}.

All measurement instruments have been psychometrically validated in several large-scale studies and show good psychometric properties\cite{Silva:2013:SPANEPortugal,Li:2013:SPANEChina,Sumi:2013:SPANEJapan,Corno:2016:SPANEItaly,Jovanovic:2015:SPANE,Benet:1998:BigFiveSpanishEng,Scheier:1994:LOTR} including consistency across full-time workers and students~\cite{Silva:2013:SPANEPortugal}.
For all questionnaires we used a further psychometrically validated German version, i.e.~\cite{Rahm:2017:SPANEGermany} for SPANE,~\cite{Lang:2001:TestguteBFI} for the Big Five Inventory and~\cite{Glaesmer:2008:LOTRdeutsch} for LOT-R.

\subsection{Hypotheses, Parameters, and Variables}
\label{methods_variables}

The independent variable relevant to RQ1 and RQ2 is the \textbf{displayed metric value (DMV)}.
We developed the DMV to express the understandability of the source code. The DMV ranges from 0 (very easy to understand) to 10 (very hard to understand). The choice was dictated to how natural it is for human beings to rate a concept from 0 to 10. The individual participant only saw three values for the metric. For the two introductory examples to explain the study tasks, we have chosen the values 1 (a very easy task) and 9 (a very hard task) to show possible extremes for coding snippet understandability. For all three experiment tasks, a participant either saw a value of 4 (moderately easy) or 8 (moderately hard), to cause the anchoring effect into two opposed directions (easy and hard).

Regarding RQ1, the relevant dependent variable is \textbf{perceived understandability (PU)}.
Perceived understandability is defined as the sum of a participant's ratings for the three code snippets.
The rating of each code snippet ranges, identical to the DMV, from 0 (very easy to understand) to 10 (very hard to understand).
Accordingly, the value for PU is in the range of 0 to 30.\\

\noindent
$H1_{0}$: There is no significant difference in perceived understandability ($PU$) between the two anchoring directions of a displayed metric value ($DMV$).\\
$H1_{A}$: There is a significant difference in perceived understandability ($PU$) between the two anchoring directions of a displayed metric value ($DMV$).\\

Regarding RQ2, we consider two common measures for code comprehension, which are time needed to complete all three tasks and correctness of the answers to the comprehension questions.
The time was recorded for each task and summed up at the end.
Correctness is the sum of correct answers to the comprehension questions of all three tasks, including both correct answers to actual return values and correct answers to return values according to the documentation.
Therefore, the value for correctness is in the range of 0 to 30.
To answer the research question, we combine time and correctness, as Scalabrino et al.~\cite{Scalabrino:2019:Automatically} did, for example, to score participants higher that are both fast and correct.
This results in \textbf{timed actual understanding (TAU)}, a participant's performance score obtained by combining correctness and time.
Equation \eqref{eq:TAU} provides the calculation for TAU, which ranges from 0 (the worst possible) to 1 (the best possible) and in which $t_{max}$ is the time of the participant who took the longest.

\begin{equation}
\frac{correctness}{30} * \left(1 - \frac{time}{t_{max}}\right)\label{eq:TAU}
\end{equation}

\noindent
$H2_{0}$: There is no significant difference in timed actual understanding ($TAU$) between the two anchoring directions of a displayed metric value ($DMV$).\\
$H2_{A}$: There is a significant difference in timed actual understanding ($TAU$) between the two anchoring directions of a displayed metric value ($DMV$).\\

The investigation of RQ3, the extent to which individual characteristics influence the deviation from the displayed metric value, is exploratory research.
Therefore, no hypotheses were formulated for this research question.
The \textbf{metric deviation} is defined as mean absolute deviation of a participant's rating from the displayed metric value and calculated as shown in \eqref{eq:MD}, where $PU_{i}$ is the perceived understandability for task i.

\begin{equation}
\frac{|PU_{1} - DMV| + |PU_{2} - DMV| + |PU_{3} - DMV|}{3}\label{eq:MD}
\end{equation}

The individual characteristics of interest in this study are experience with Java, personality, happiness and dispositional optimism and pessimism.
Java experience was measured in years.
Personality was operationalized by the five dimensions of the Five Factor model, i.e., extraversion $range=[0,32]$, agreeableness $r=[0,36]$, conscientiousness $r=[0,36]$, neuroticism $r=[0,32]$ and openness to experience $r=[0,40]$.
The higher the value, the more pronounced is the respective personality facet of a participant.
The range of SPANE-P (positive affect) and SPANE-N (negative affect) is $r=[6,30]$, from low frequency to high frequency of positive and negative experiences, respectively.
SPANE-B, or happiness, has a range $r=[-24,24]$, the negative pole refers to unhappiness and the positive one to happiness.
Dispositional optimism and pessimism are independent constructs rather than a bipolar trait.
Both are in the range $r=[0,12]$, from low to high degree of optimism and pessimism, respectively.

\subsection{Experiment Design}
\label{methods_design}

The experiment was a between-subject design with two treatment groups. Assignment to a treatment was double-blind and random.\footnote{We agree, to some extent, with Baltes and Ralph~\cite{Baltes:2020:Sampling} that random should be used sparingly, so we will add here that participants were assigned to a condition based on the time slot they signed up for. When assigning treatment conditions to a time slot, it was ensured that the conditions were distributed equally over different times of day.} None of the authors knew which participant, even as anonymous data point, was in which treatment group until the data was evaluated.

One group saw a $DMV$ of 4 next to all code snippets. We call this group the \textit{easy} group from this point on. The other group saw a $DMV$ of 8. We refer to this group as the \textit{hard} group from this point on. There were no further differences in the treatment of the two groups.

The reader might notice an absence of a control group, which does not see any $DMV$. This is by design and suggested in literature on the anchoring effect, which we discuss in more detail in~\ref{sec:limitations}.

Following Wohlin et al.~\cite{Wohlin:2012:Experimentation} guidelines for conducting experiments in software engineering, we had the study design reviewed by two peers in two iterations and conducted a pilot test.
Furthermore, we have identified and implemented a number of measures that we believe contribute to mitigating threats to validity. Limitations of our final study design, that is, what affects the interpretation of our results, are discussed in~\ref{sec:limitations}.

\subsubsection{Measures to Address Construct Validity}

We used psychometrically validated questionnaires for assessing individual characteristics.
Regarding the operationalization of code understanding, we have oriented ourselves on how the construct was measured by our peers in peer reviewed research~\cite{Scalabrino:2019:Automatically, Baron:2020:CogComplexity}.
With time and correctness we measured two argumentatively important and objective aspects of code understanding and combined them with equal weight similar to what Scalabrino et al. did~\cite{Scalabrino:2019:Automatically}.

We designed a plausible scenario to justify the display of the metric value and developed an experiment description for participants which did not reveal the essence of the experiment to prevent \textit{hypothesis guessing}~\cite{Wohlin:2012:Experimentation}. The description did not present false information, but hid only the objective of the anchoring effect.
Closely related, we prevented the threat of \textit{experimenter expectations}~\cite{Wohlin:2012:Experimentation} by double-blind assignment of participants to treatment groups.

To prevent \textit{evaluation apprehension}~\cite{Wohlin:2012:Experimentation} and unnecessary stress we always tried to have exactly two participants simultaneously in the room, the experimental supervisor could not look over the shoulder of the participants and it was emphasized several times that the answers were anonymous.
Due to their intimate nature, the Big Five and LOT-R questionnaires were completed only after the code comprehension tasks had been completed.
Both participants in the same time slot were also in the same treatment group to avoid \textit{diffusion or imitation of treatments}~\cite{Wohlin:2012:Experimentation} for example by one of the two participants making a comment on the displayed metric value.

\subsubsection{Measures to Address Internal Validity}

We discussed every variable listed in the mapping study on confounding variables in code comprehension experiments~\cite{Siegmund:2015:Confounding}.
Of the 37 variables listed, only two remained even after thorough planning of the experiment, which we see as potential threats to validity: the Hawthorne effect and selection (the generalizability of student participants).
We discuss both in~\ref{sec:limitations}.

Of the potential confounders that we have explicitly controlled, we highlight the following two.
First, we selected several code snippets that a participant had to understand to reduce the influence of individual data structures and program semantics.
Snippets were also neither too difficult nor too easy according to a validated code comprehensibility metric~\cite{Baron:2020:CogComplexity}, and moreover of comparable comprehensibility.
Second, we implemented a tool to view and interact with the code in the tasks.
This gave us full control over the environment and allowed us to reduce the displayed elements to a minimum to increase internal validity.
In addition, no participant had an advantage, since no one was more familiar with the environment than everyone else.

\subsubsection{Measures to Address External Validity}

We used code snippets taken from real-world, actively developed famous open-source projects and avoided removing comments or obscuring method or variable names to force code understanding.
Instead, we have created a realistic software maintenance scenario in which the developer needs to understand code that does not necessarily fit the documentation and that may contain a bug, requiring the developer to check what the actual output values are for certain input values.

\subsubsection{Measures to Address Conclusion Validity}

Regarding \textit{reliability of treatment implementation}~\cite{Wohlin:2012:Experimentation}, we used a strict protocol and double-blind condition assignment to ensure that each participant received the same information and was treated equally.
The same investigator conducted the experiment with all 45 participants.
The $DMV$ of either 4 or 8 was displayed the same for all participants in a treatment group, as it was an automatic display in an environment controlled by us.

To prevent \textit{random irrelevancies in experimental setting}~\cite{Wohlin:2012:Experimentation} we reserved the room two weeks in advance and for the entire day on each day the study was conducted.
We were prepared to document irregularities, but did not have to do so.

\subsection{Procedure}
\label{methods_procedure}

From the moment one or two participants arrived in the room reserved for the experiment at the agreed time, the following steps took place.
Participants were provided with a consent form and were informed verbally about the aim of the study.
They were shown the laptop and the files on the laptop.
Then they had to fill out two questionnaires, one on demographic data and on their happiness (SPANE).
The instructor made the participants save and close the questionnaires on their own after completing them, so that they did not feel observed by the instructor.

The instructor explained the task, the scenario and the development environment to the participants.
Thereby the participants were shown a filled out task sheet for one method as an example and the solution was exemplified for the first two input values.
The visualization of the metric and its display for supposedly informative reasons was explained and a second introductory example, which was significantly harder to understand than the first one, demonstrated that the metric works well.
When asked, the instructor told that the metric was developed by experts, similar to what~\cite{Draganich:2014:PlaceboSleep} did in their placebo sleep experiment.

The instructor summarized the task and mentioned that the employer in the task scenario wants them to work efficiently but also correctly.
Participants then had the opportunity to ask questions before starting with the tasks.
The time recording was done per task and the recorded value was stored in a spreadsheet without the participants noticing it.
After all three tasks had been completed, the participants finally had to fill out the questionnaires on individual characteristics.

%
%

\section{Results}
\begin{figure}[t]
\centering
\includegraphics[width=\columnwidth]{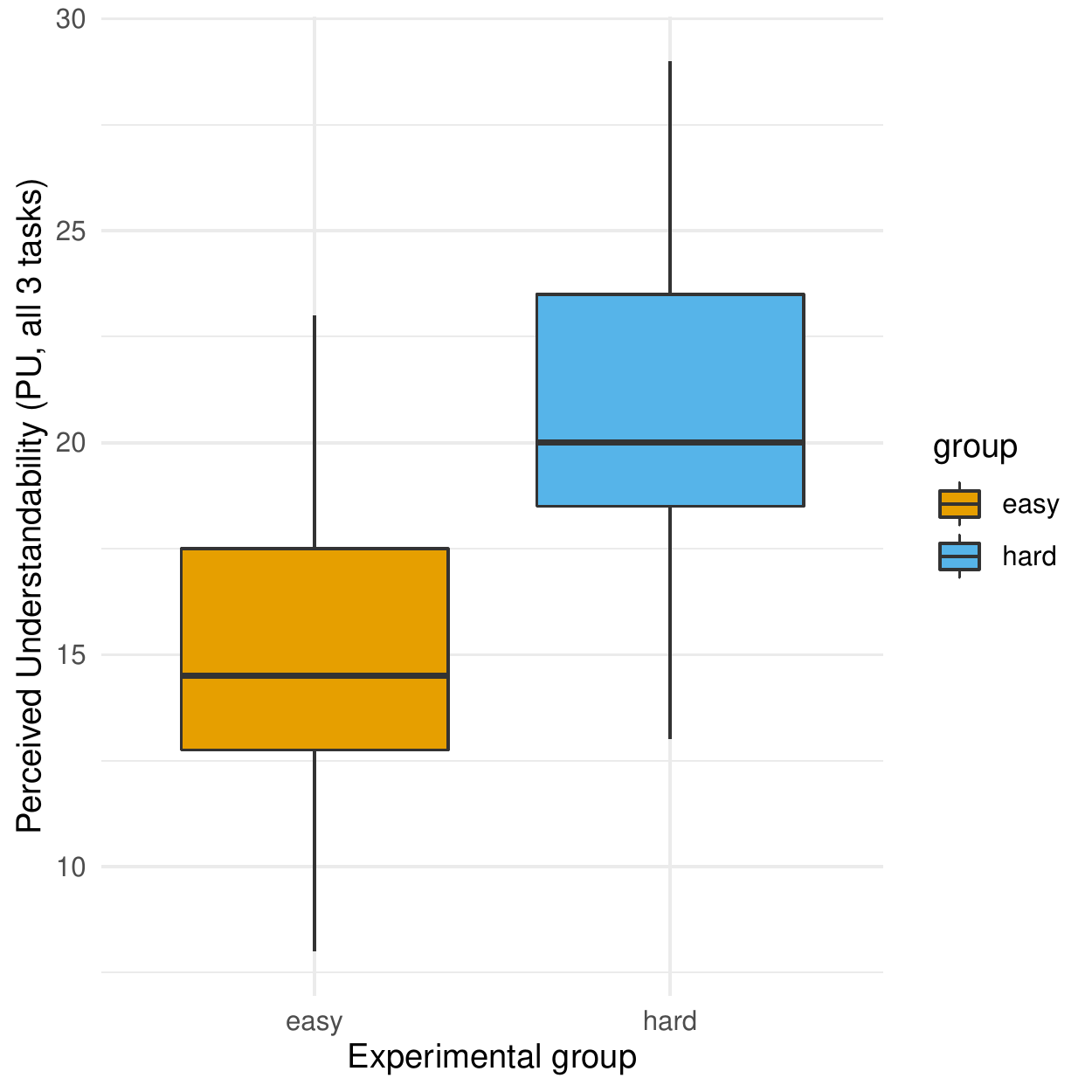}
\caption{Perceived understandability (PU) of the tasks for the easy group and the hard group (three tasks, range 0 (easiest) to 10 (hardest), combined range 0 (easiest) to 30 (hardest)).}
\label{fig:boxplot-rq1}
\end{figure}

\begin{table*}[t]
\caption{Spearman's $\rho$ correlation coefficient of individual characteristics with the deviation of the subjective rating.}
\centering
\resizebox{\textwidth}{!}{%
\begin{tabular}{rrrrrrrrrrrrr}
\hline
Group & Java Experience (years) & SPANE-P & SPANE-N & SPANE-B & Optimism & Pessimism & Extraversion & Agreeableness & Conscientiousness & Neuroticism & Openness \\
\hline
easy & -0.28 & 0.11 & -0.23 & 0.18 & 0.32 & 0.18 & -0.1 & 0.08 & 0.46 & -0.12 & -0.01 \\
hard & 0.24 & -0.15 & 0.13 & -0.23 & 0 & -0.04 & -0.09 & 0 & -0.09 & -0.06 & 0.05 \\
combined & -0.03 & -0.12 & 0.06 & -0.16 & 0.14 & 0.07 & -0.13 & 0.08 & 0.14 & -0.05 & 0.01 \\
\hline
\end{tabular}
}
\label{tab:rq3}
\end{table*}

\subsection{Descriptive statistics and dataset preparation}
45 students participated in the study\footnote{We cannot offer a precise acceptance rate because course attendance is not mandatory and cannot be recorded.}. Two of them did not submit complete data for RQ3 (e.g., missing an item for the SPANE questionnaire). We decided to exclude them from the dataset to enhance our confidence in how serious all participants were in completing all tasks. We thus had an overall sample size of $n = 43$ participants ($41$ male, $2$ female). Mean age was $24.47$ ($SD =2.84$), average declared experience with the Java programming language was $5.83$ years ($SD =2.37$).

Participants were randomly allocated to the \textit{easy} group, $n = 20$) or the \textit{hard} group, $n = 23$. Declared experience with the Java programming language was comparable for both groups after random assignment ($M = 5.25$, $SD =2.60$ for the easy group, $M = 6.33$, $SD =2.06$ for the hard group).

The easy group was shown a $DMV$ of $4$ for the three tasks, or $12$ combined, and provided a $PU$ of $15.40$ ($SD =4.17$, $median = 14.50$). The hard group was shown a $DMV$ of $8$ for the three tasks, or $24$ combined, and provided $PU$ of $20.83$ ($SD =4.23$, $median = 20.00$). A graphical comparison is offered in the boxplot of fig.~\ref{fig:boxplot-rq1}.

The easy group performed with an average $TAU$ of $M = .37$ ($SD =.17$, $median = .41$).
The hard group performed with an average $TAU$ of $M = .37$ ($SD =.11$, $median = .36$). A further boxplot comparison (included in the supplemental material) did not suggest significant difference.

\subsection{Hypothesis testing}

There was a significant difference between $PU$ of the two groups\footnote{Welch Two Sample t-test given evidence for non-normality, (Shapiro-Wilk test, $p > .26$ for both groups) and no further assumption on the population variance.}\footnote{We are aware of the open debate on whether Likert items are ordinal data or continuous data~\cite{murray2013likert}. We believe, in line with psychometric theory, to have Likert items capture discrete points over a continuous scale. All scales that we use for individual characteristics are psychometrically validated Likert items.}, $t(40.318) = -4.227$, $p = .000132$, 95\% CI $[-8.02, -2.83]$, with a large effect size, $d = -1.29$, 95\% CI $[-1.97, -0.61]$. 

We thus reject $H1_0$ in favor to $H1_A$: \textbf{There is evidence for a difference in perceived understandability between the two anchoring directions of a displayed metric value}.

There was no significant difference between the timed actual understanding ($TAU$) of the two groups\footnote{Wilcoxon rank sum exact test given evidence for non-normality (Shapiro-Wilk test, $p = .03$ for the easy group).}, $W = 256$, $p = .5385$, with a negligible effect size, $d = -.006$, 95\% CI $[-0.62, 0.61]$. 

We do not reject $H2_0$. \textbf{There is no evidence for a difference in timed actual understanding between the two anchoring directions of a displayed metric value}.

\subsection{Exploratory analysis}

We provide in Table~\ref{tab:rq3} the computed correlation coefficients of the metric deviation (calculated as in formula~\ref{eq:MD}) with affect-related metrics and personality-related metrics.
The first two rows provide the correlation coefficients for the two experimental groups (between), while the third row combines all participants (within).

For brevity's sake, we will call the ``metric deviation'' simply ``deviation'' the rest of this section. As a reminder, the wider the deviation, the bigger the gap between the subjective rating and the shown metric value, or the manipulation, on the easy direction or on the hard direction.

Given the exploratory nature of RQ3, no estimation of significance was conducted for the correlation coefficients. As a cutoff for potentially interesting individual characteristics, we will only consider correlation coefficients $|\rho| > 0.1$.

When exploring the data between the two experimental groups, we notice that an increase in programming language experience is associated with a decreased deviation when the manipulation suggests an easy task and an increased deviation when the manipulation suggests a hard task. The opposite happens with happiness (SPANE-B). An increase in happiness is associated with an increased deviation for the easy group and a decreased deviation for the hard group. The two major components of happiness, SPANE-P and SPANE-N, show coherence with the aggregated happiness score.

An increase for \textit{both} optimism and pessimism is associated with a wider deviation for the easy group, while no correlation is observed for the hard group.
Of all personality traits, an increase in conscientiousness seems to be strongly correlated with an increased deviation for the easy group, followed by neuroticism but with an inverse relationship. No personality trait shows a $|\rho| > 0.1$ for the hard group.

When combining the two groups, in a within subject analysis, with the assumption that the two groups are from the same population, an overall negative relationship between happiness and the deviation is observed (happier participants deviated less). An increase in optimism is associated with a wider deviation. Of the personality traits, an increase in optimism and conscientiousness were correlated with a bigger deviation, but an increase with extraversion was correlated with smaller deviation. Programming experience seems to not play a role in the deviation.

%
%

\section{Discussion}
\label{sec:discussion}

Metrics provide developers with quantitative insights into the quality of their source code, but most of the metrics used in practice are not sufficiently validated.
We have investigated the extent to which developers are actually subconsciously influenced by the value of a displayed made-up metric to create an awareness of responsibility in code quality reporting.

We found a significant and strong anchoring effect, which means that developers are strongly influenced by a displayed metric value in their rating of source code comprehensibility.
This finding is consistent with over 40 years of research on the anchoring effect~\cite{Furnham:2011:ReviewAnchoring}, yet investigation in a specific software engineering context is nevertheless a valuable contribution to build the foundation for future studies, for example to investigate consequences of the demonstrated effect.

One such potential consequence could be an improved or worsened understanding of the code.
However, in our study we could not provide evidence that the suggestion of simple or difficult code provokes a placebo effect, in the sense that the beliefs concerning the comprehensibility of the source code influences the speed and correctness with which a software engineer answers comprehension questions.
Since cognitive performance and creativity, arguably both characteristics necessary for code understanding, can be influenced by a placebo~\cite{Turi:2018:CognitivePlacebo,Rozenkrantz:2017:Creativity}, we would have expected to observe such an effect.
Assuming that a placebo effect for code comprehension can be observed theoretically, we see two possible reasons why we could not in our experiment.

First, it could be that a displayed metric value is simply not a strong enough influence to cause performance changes.
Compared to other placebo studies~\cite{Draganich:2014:PlaceboSleep,Crum:2007:Mindset,Rozenkrantz:2017:Creativity}, we did not manipulate the participants to claim that our treatment had specific beneficial or performance-enhancing effects.
Instead, we displayed a metric value that indicated how easy or difficult code is to understand.
We left it up to the participants to interpret what it means and what consequences it might have if source code is easy or hard to understand.
Accordingly, code comprehension may have actually improved or worsened as a result of the manipulation, just not in the way we measured it.

This brings us to the second possibility that time and correctness are not all-inclusive proxies for code comprehension.
We speculate that physiological measurements may have led to a different result.
For example, cognitive load and stress levels at the end of task processing might have been lower in the easy group, as they may have been more comfortable with the task.
While the question of ideal measures of code comprehension is beyond the scope of this work, and we have not measured these variables, we argue that they nevertheless reflect relevant dimensions of code comprehension.
With the increase in physiological measurements in the field of code comprehension~\cite{Siegmund:2020:Crazy,Fakhoury:2018:Objective,Peitek:2018:Simultaneous}, we see much potential for future studies to replicate our experiment with alternative measurement methods to shed light on the matter.

Regarding individual characteristics that influence the strength of the observed anchoring effect, our results, while exploratory and based on correlations, are only partially consistent with the few relevant studies conducted to date~\cite{Furnham:2011:ReviewAnchoring}.
We observed that participants with low extraversion are less subjected to the anchoring effect, which is consistent with existing literature~\cite{Eroglu:2010:Biases}.
However, our results indicate that participants with high conscientiousness are less susceptible to the anchoring effect, which is contrary to the findings of Eroglu and Croxto~\cite{Eroglu:2010:Biases}.
We could not find a positive correlation between anchoring and the personality trait openness as McElroy and Dowd~\cite{Mcelroy:2007:Susceptibility} did.
Further, our results suggest that happier people might be more subjected to the anchoring effect, which also does not coincide with previous studies~\cite{Englich:2009:Moody,Bodenhausen:2000:Sadness}.

We found that optimism positively correlates with the metric deviation, which corresponds to a weaker anchoring effect.
Programming experience on the other hand might not play a role in anchoring, but this should be taken with caution, since we had a homogeneous group of experienced developers and results could be different for inexperienced developers.

If the code comprehensibility metric shows a low value for a rather difficult task, personality factors could play a bigger role.
In particular, conscientiousness might be the strongest predictor of a deviation when metrics are too conservative in how difficult a task might be (easy group), followed by optimism.
In conclusion, our results suggest that the anchoring effect might not be a universal rule that applies equally to all participants. The partial deviation of our results from previous findings is an indication that more studies are needed in this regard. The exploratory settings of RQ3 set basic building blocks upon which we call our peers to conduct future research.

\subsection{Limitations}
\label{sec:limitations}

A detailed description of design decisions made in advance to mitigate validity risks can be found in~\ref{methods_design}.
What follows are limitations that affect the final study design and should be considered when interpreting our results.

As for potential confounding variables, we could reduce a lengthy list of known variables~\cite{Siegmund:2015:Confounding} to two that might have affected the results: the Hawthorne effect and the selection of students as participants.

The Hawthorne effect~\cite{Roethlisberger:1939:Management, Mccarney:2007:Hawthorne} describes that participants in experiments would behave differently because they were observed.
While we addressed \textit{hypothesis guessing}~\cite{Wohlin:2012:Experimentation} with a plausible scenario that does not reveal the essence of the experiment, and we addressed the threat of \textit{evaluation apprehension}~\cite{Wohlin:2012:Experimentation} through privacy and data anonymization, it cannot be ruled out that participants followed the proposed metric value more closely than they would have done outside an experimental setting.
According to~\cite{Furnham:2011:ReviewAnchoring} the current dominant view of the anchoring paradigm focuses on confirmatory hypothesis testing in the sense that information is activated that is consistent with the anchor presented.
We assume that this also applies to our experiment and that the observed anchoring effect, if at all, is only to a small extent due to an experimentally provoked good will of the participants for the metric.
A future field study could provide clarity in this respect.

As argued earlier in the description of participants~(\ref{methods_participants}), we invited a convenience sample of students of a software engineering MSc study program. In our sampling strategy, we prioritized internal validity, which was enhanced by a homogeneous level of experience with the programming language, paradigm, and task type. We also believe that our results can be generalized to a population of professional software engineers. In the discussion about representativeness of software engineering students, opposing opinions have existed for the last 20 years~\cite{Kitchenham:2002:Preliminary,Sjoberg:2002:Realistic,Tichy:2000:Hints}.
Recent studies have shown that comparable results can be achieved with both groups of students and professionals---as long as the scope of the investigation is carefully considered (see, e.g.,~\cite{Salman:2015:students}). We have confidence in the robustness and soundness of our research design. 
Recent commentaries~\cite{Feldt:2018:FourCommentaries} have, once again, highlighted how diverse the views are on the topic.
We side with the view summarized by Runeson~\cite{Feldt:2018:FourCommentaries} as well as Baltes and Ralph~\cite{Baltes:2020:Sampling} that a convenience sample of students is justified in the investigation of central behavioral and cognitive processes, as was the case in our study. There is evidence, for example, that the anchoring effect is not restricted to laymen and that more experienced people are influenced by it as well~\cite{Tversky:1974:Anchoring, Furnham:2011:ReviewAnchoring}.

Then, since we did not have a control group in our experiment that was \textit{not} shown a metric value, we cannot say how a control group would have rated the snippets.
For the demonstration of the anchoring effect this is not a limitation and it is consistent with the body of research on the anchoring effect to not have a no-anchor control group~\cite{Critcher:2008:Incidental,Mussweiler:2005:Subliminal}.
However, regarding the placebo effect, we would like to stress that our design would not be able to decide strictly speaking whether any observed effect is due to the placebo, but with the design it is still possible to demonstrate the extent to which a displayed metric value influences code understanding in a positive or negative direction.
Our study is based on a comprehensive body of research that has provided evidence for the placebo effect, so we assumed that the effect would also exist in our scenario and opted for the study design described above.

Finally, we are aware that the way in which the metric value was displayed is very prominent.
We argue that developers are used to static code analysis tools reporting metric values in similar ways and IDEs increasingly offer the possibility to display code quality metrics directly in the source code.
Even if this situation is not as common, we refer to Critcher and Gilovich~\cite{Critcher:2008:Incidental} and related works showing that for the anchoring effect to work, much more inconspicuous anchors, which do not even have to be highlighted, are usually sufficient.
Again, a field study with realistic IDE plugins and existing metrics could be an option to repeat the investigation of the effects in an industrial environment.

\subsection{Implications}

Since developers are influenced by a shown metric value in their subjective evaluation of a code snippet's comprehensibility, we highlight the following implications.

First, those responsible for reporting code quality metrics should be aware of their responsibility that non-validated metrics lead to unwarranted manipulation of developers, the consequences of which we do not know yet.
Future studies can build on our results and investigate possible consequences.

Second, since it is a common practice in code understanding studies to ask developers for a subjective rating, such studies should ensure that individual participants are not anchored by context factors such as displayed metric values.
It may already be sufficient that the instructor or the task description hint at something about the complexity of a code snippet to be examined.
Also, for example, different amounts of time available for processing different code snippets could lead to an anchoring of the participants in their subjective assessment.
If the study cannot be controlled with certainty in this regard, the measure of subjective ratings should not be used.

We echo the call of Mohanani et al.~\cite{Mohanani:2018:Cognitive} and propose a debiasing technique for the anchoring effect. The debiasing here is about developing validated metrics (or validate existing ones) before showing their values to software developers.

We do not consider it an issue when developers are anchored in their subjective judgement by a validated metric that may be able to consider more factors and evaluate more objectively than a developer could.
We are aware that, for example, static code analysis tools do not intentionally lie, and that many of the metrics seem to make intuitive sense.
It becomes problematic, however, when developers interpret quality aspects into metrics that were not intended to be measured by the metric, or when the metric is no more than an implemented, albeit well-thought-out, idea.
Therefore, tools should describe very precisely what the metric intends to measure and support this measurement with systematic research.

A number of previous studies already called for more effort to be put into disseminating research findings among practitioners so that they can rely on evidence rather than forming biased and error-prone conclusions based on personal impressions~\cite{Devanbu:2016:Belief,Kitchenham:2004:Evidence}.
Especially with respect to the clearly shown anchoring effect, validated metrics should have an easy time anchoring developers where they should be anchored evidence-wise and overcome the circumstance that developers sometimes tend to prefer their own opinions over empirical evidence~\cite{Rainer:2003:Persuading}.

We consider the negative results on RQ2 to be good results under the circumstances described above.
Since the situation is unlikely to change in the near future, it is at least good to know that a few random numbers may not have a negative impact on a developer's understanding time and correctness during maintenance.
Whether other aspects of code understanding can be influenced by a placebo will need to be investigated in future studies.

%
%

\section{Conclusion}

We investigated whether the value of a shown code comprehensibility metric influences subjective ratings of code comprehensibility and actual code comprehension performance, i.e., the time spent and correctness in answering comprehension questions.
In a randomized double-blind experiment two groups of participants had to understand three code snippets, answer comprehension questions and rate the understandability of the code snippets.
We found the shown metric value to have a significant and large effect on the developers' ratings but not on their performance.
The strength of the anchoring effect appears smaller for optimistic and conscientious people.

Since we have limited understanding of the consequences of the demonstrated manipulation of developers by non-validated metrics, we call for an increased awareness of the responsibility in code quality reporting and for corresponding tools to be based more strongly on scientific evidence.
Studies in which code comprehensibility is measured via subjective ratings should control contextual factors such as shown metric values that may potentially influence developers' ratings, or refrain from this measure altogether.

Future works should focus on investigating the consequences of the belief that code is easy or hard to understand.
A similar study should be conducted to investigate the influence of displayed metric values on other dimensions of code understanding, such as cognitive load, and we consider a slightly modified version of our study to be useful, e.g., to investigate the influence of code complexity on the strength of the anchoring effect.

\section{Data Availability}

Following open science principles in software engineering~\cite{mendez2020open}, we disclose code snippets, task sheets with comprehension questions, anonymized raw data, and the R script for the analysis openly~\cite{zenodo:dataset}.

\section*{Acknowledgments}
We are deeply thankful to our participants for taking part in our study. We thank three anonymous reviewers for their insightful comments and Katharina Plett for proofreading.

\bibliographystyle{IEEEtran}
\bibliography{bibliography}

\end{document}